\documentclass[aip,
 mph,%
 amsmath,
 amssymb,
 superscriptaddress,
 floatfix,
 preprint,
 eqsecnum,%
 numerical,%
]{revtex4-1}

\usepackage{graphicx}
\DeclareGraphicsExtensions{.pdf,.png,.jpg}
\usepackage[caption=false%,labelformat=parens,subrefformat=subparens,listofformat=subparens
]{subfig}
\usepackage{epsfig}
\usepackage{epstopdf}
\usepackage{lipsum}
\usepackage{wrapfig}
\usepackage{wasysym}
\usepackage{threeparttable,booktabs,siunitx}
\usepackage{booktabs}
\usepackage{lmodern}
\usepackage{longtable}
\usepackage{multirow}
\usepackage{amsmath}
\usepackage[OT2,T1]{fontenc}
\usepackage[bookmarks=false]{hyperref}
\usepackage{color}
\hypersetup{pdfauthor={},pdfborder=0 0 0,colorlinks=true,citecolor=blue,linkcolor=blue,urlcolor=blue,}

\DeclareSymbolFont{cyrletters}{OT2}{wncyr}{m}{n}
\DeclareMathSymbol{\comb}{\mathalpha}{cyrletters}{"58}

\newcommand{\ben}{\begin{eqnarray}\displaystyle}
\newcommand{\een}{\end{eqnarray}}

\begin{document}

\title{Estimation of angular sensitivity for X-ray interferometers with multiple phase gratings}
% Force line breaks with \\

\author{Jianwei Chen}
\affiliation{Research Center for Medical Artificial Intelligence, Shenzhen Institutes of Advanced Technology, Chinese Academy of Sciences, Shenzhen, Guangdong 518055, China.}%Lines break automatically or can be forced with \\
\author{Jiecheng Yang}%
 \affiliation{Research Center for Medical Artificial Intelligence, Shenzhen Institutes of Advanced Technology, Chinese Academy of Sciences, Shenzhen, Guangdong 518055, China.}
\author{Peiping Zhu}
\affiliation{University of Chinese Academy of Sciences, Beijing, 100049, China.}%
\affiliation{Institute of High Energy Physics, Chinese Academy of Sciences, Beijing, 100049, China.}%
\author{Ting Su}
\affiliation{Research Center for Medical Artificial Intelligence, Shenzhen Institutes of Advanced Technology, Chinese Academy of Sciences, Shenzhen, Guangdong 518055, China.}%Lines break automatically or can be forced with \\
\author{Huitao Zhang}
\affiliation{School of Mathematical Sciences, Capital Normal University, Beijing, 100048, China.}
\author{Hairong Zheng}%
\affiliation{Paul C Lauterbur Research Center for Biomedical Imaging, Shenzhen Institutes of Advanced Technology, Chinese Academy of Sciences, Shenzhen, Guangdong 518055, China.}%
\author{Dong Liang}%
 \email{dong.liang@siat.ac.cn.}
 \affiliation{Research Center for Medical Artificial Intelligence, Shenzhen Institutes of Advanced Technology, Chinese Academy of Sciences, Shenzhen, Guangdong 518055, China.}
\affiliation{Paul C Lauterbur Research Center for Biomedical Imaging, Shenzhen Institutes of Advanced Technology, Chinese Academy of Sciences, Shenzhen, Guangdong 518055, China.}%
\affiliation{University of Chinese Academy of Sciences, Beijing, 100049, China.}%
\author{Yongshuai Ge}%
 \email{ys.ge@siat.ac.cn.}
 \affiliation{Research Center for Medical Artificial Intelligence, Shenzhen Institutes of Advanced Technology, Chinese Academy of Sciences, Shenzhen, Guangdong 518055, China.}
\affiliation{Paul C Lauterbur Research Center for Biomedical Imaging, Shenzhen Institutes of Advanced Technology, Chinese Academy of Sciences, Shenzhen, Guangdong 518055, China.}%
\affiliation{University of Chinese Academy of Sciences, Beijing, 100049, China.}%

\date{\today}% It is always \today, today,

\begin{abstract}
Recently, X-ray interferometers with more than one phase grating have been developed for differential phase contrast (DPC) imaging. In this study, a novel framework is developed to predict such interferometers' angular sensitivity responses (the minimum detectable refraction angle). Experiments are performed on the dual and triple phase grating interferometers, separately. Measurements show strong consistency with the predicted sensitivity values. Using this new approach, the DPC imaging performance of X-ray interferometers with multiple phase gratings can be further optimized for future biomedical applications.
\end{abstract}

\maketitle

As one of the many X-ray phase contrast imaging techniques, the grating-based DPC imaging interferometer has attracted a lot of attention due to its high compatibility with the conventional laboratory X-ray tube and flat panel detector assembly. Over the past two decades, the so called Talbot and Talbot-Lau interferometers\cite{Momose2003, Weitkamp2005, Pfeiffer2006a} have been widely investigated, and some prototype biomedical imaging systems have already been developed. For instance, the lung imaging system\cite{sauter2019optimization}, the mammography system\cite{coan2013phase}, and so on. When designing these systems, two critical factors need to be taken into account to obtain the best DPC imaging performance: one is the angular sensitivity of the interferometer, and the other is the number of X-ray photons falling into the detector plane. The importance of angular sensitivity relies on the fact that it reflects the minimum detectable refraction angle of a certain interferometer system. Thus, its value is always expected to be high from the signal detection prospective. Meanwhile, the latter factor determines the noise level of the DPC image, and thus is always expected to be low. Aimed at achieving either super high sensitivity or better X-ray radiation dose efficiency, recently, novel X-ray interferometers with more than one phase grating have been investigate experimentally\cite{Miao2016a, kagias2017dual}. By using the transparent (almost no absorption of X-rays) phase gratings, it is possible to improve the radiation dose efficiency by a factor of about two (compared with the Talbot-Lau system with an analyzer grating). However, so far it is still not very clear how to predict and optimize the DPC signal strength (i.e., angular sensitivity) of such novel interferometers in theory yet, especially for the ones containing three phase gratings.

Mathematically, the angular sensitivity of an X-ray DPC imaging interferometer is defined via the following equation:
\begin{align}
\label{eq:sens}
\phi = S\times\varphi = S\times\frac{\lambda}{2\pi}\frac{\partial \Phi}{\partial x}.
\end{align}
In Eq.~(\ref{eq:sens}), $S$ denotes the angular sensitivity, $\lambda$ denotes the X-ray beam wavelength, the term $\varphi=\frac{\lambda}{2\pi}\frac{\partial \Phi}{\partial x}$ represents the tiny X-ray beam refraction induced by the object along $x$-axis, and $\phi$ corresponds to the measured DPC signal from the interferometer diffraction fringes.

The work\cite{yan2019sample} done by Yan \emph{et. al.} has offered one possible solution to estimate the angular sensitivity of the dual phase grating interferometer system. In their analyses\cite{yan2018quantitative}, it is assumed that the final diffraction fringes are formed due to the beating effect between the fringes generated by the two phase gratings. Therefore, the sensitivity of the dual phase grating interferometer was estimated in an overlapping way, especially when estimating the sensitivity behind the second phase grating. More specifically, the sensitivity behind the second phase grating has two contributors: one comes from the first phase grating G1, and the other comes from the second phase grating G2. Despite of its feasibility, it is obvious that such estimation method for three phase grating system would become much more complicated. In addition, the sensitivity between the two phase grating is not explicitly provided in that work.

\begin{figure}[t]
\centering
\includegraphics[width=0.85\textwidth]{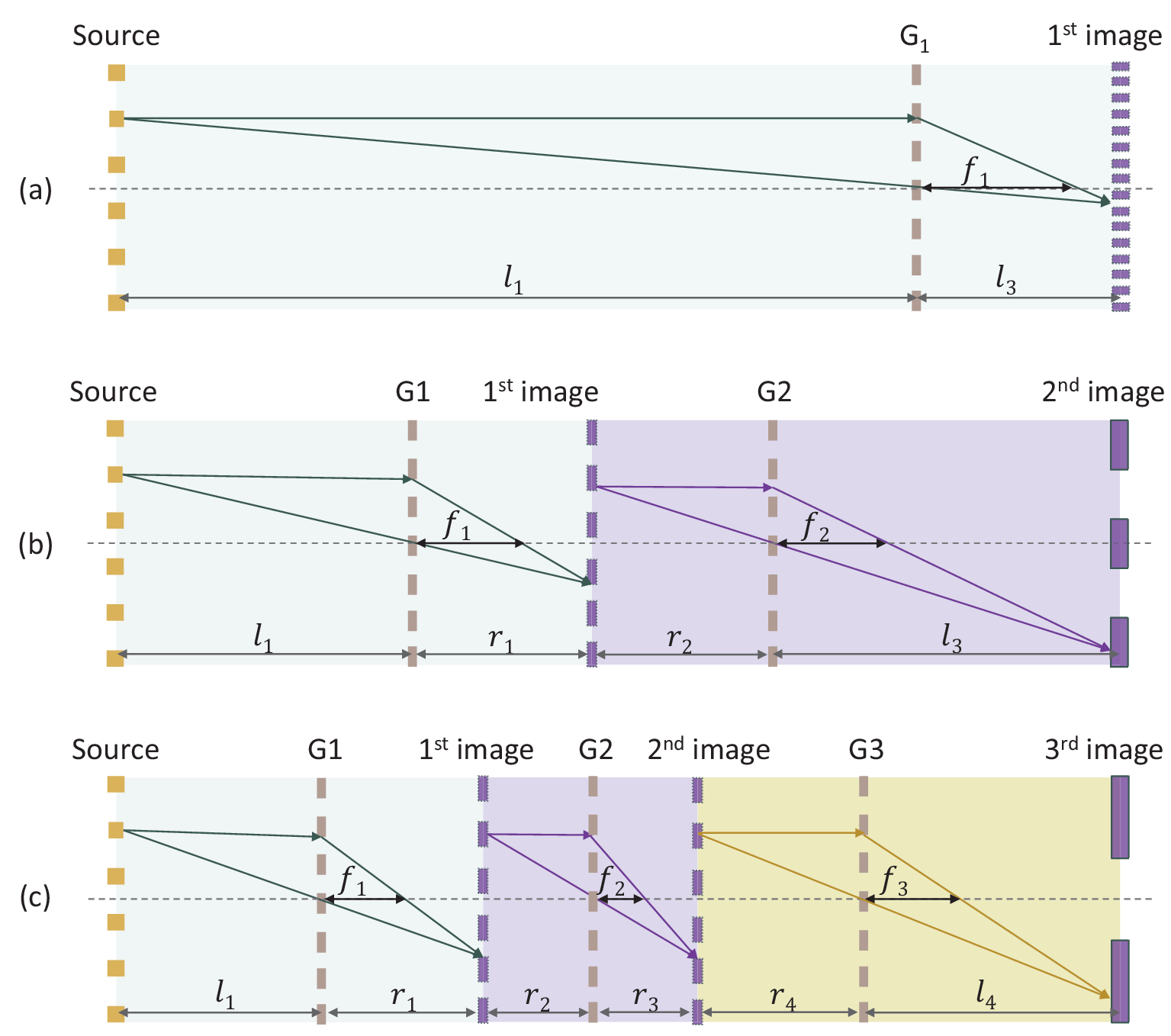}
\vspace{-0.1in}
\caption{Schemes of different X-ray interferometers with: (a) one phase grating, (b) two phase gratings, and (c) three phase gratings. The X-ray optical modules (XOM) are highlighted with certain colors (color available online). The focal length $f_i$ of the $i$-th phase grating, and the corresponding image formed after that grating are all depicted. An arrayed source is assumed, and no object is added.}
\label{fig:fig1}
\end{figure}

The aim of this study is to establish a new general framework to estimate the sensitivity of X-ray interferometers which contain any number of phase gratings with an object being inserted at any position between the source and the detector. The core of this new method is based on our thin lens imaging theory developed for the dual phase grating interferometer\cite{ge2020dual}. In the previous theory, the role of an X-ray phase grating was considered as a thin optical lens. As a result, the formed diffraction fringe with a large period in the dual phase grating interferometer can be explained as a magnified image of the source image (formed between the two phase gratings), see Fig.~\ref{fig:fig1}(b). Essentially, the entire imaging procedure is separated into two cascaded stages: the first stage is related with the G1 phase grating imaging of the X-ray source; the second stage is related with the G2 phase grating imaging of the fringe image formed by the G1 phase grating. Obviously, our explanation to the dual phase grating DPC imaging procedure is different to the theory offered by Yan \emph{et. al}. Therefore, it is necessary to provide a new angular sensitivity estimation framework for the X-ray interferometers having multiple phase gratings. In this study, we mainly focus on discussing the dual phase grating interferometer and the triple phase grating interferometer, see Fig.~\ref{fig:fig1}(b)-(c).

\begin{figure}[t]
\centering
\vspace{-0.10in}
\includegraphics[width=0.75\textwidth]{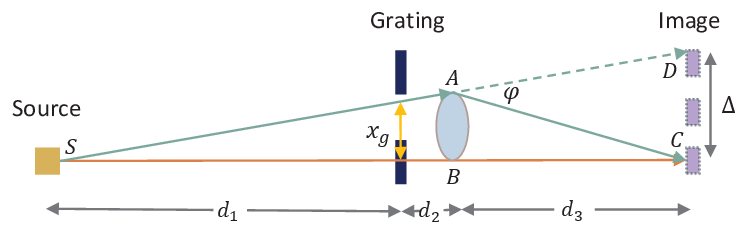}
\vspace{-0.1in}
\caption{Illustration of the object-induced X-ray refraction with one single XOM. Dashed ray AD shows the primary beam which is refracted to beam AC at the object edge. The object-induced refraction angle is denoted as $\varphi$. On the fringe image plane, the corresponding diffraction fringe shift is denoted as $\Delta$.}
\label{fig:fig2}
\end{figure}

Intuitively, the sensitivity of different grating interferometers, either the conventional Talbot or Talbot-Lau interferometer, or the ones containing multiple phase gratings, should share the similar expressions and interpretation mechanisms in physics. Motivated by this idea, an X-ray optical module (XOM) is assumed to facilitate the following discussions. By definition, an XOM contains three components: the illumination source, the phase grating, and the image of the source (diffraction fringe). Be aware that the illumination source could be either a real X-ray tube source or the diffraction fringe formed behind the phase grating. Thus, it is easy to see that the conventional Talbot-Lau interferometer only contains one single XOM, see Fig.~\ref{fig:fig1}(a). Moreover, there are two cascaded XOMs in the dual phase grating interferometer system, see Fig.~\ref{fig:fig1}(b). Finally, there are three cascaded XOMs in a triple phase grating interferometer system, see Fig.~\ref{fig:fig1}(c).

As illustrated in Fig.~\ref{fig:fig2}, assuming the object is positioned between the grating and the diffraction image of the source, the refraction of primary X-ray beam AD at the object edge (point A) causes a certain diffraction fringe shift, denoted as $\Delta$, from point D to point C on the detector plane\cite{donath2009inverse}. According to the well-known phase stepping (PS) model\cite{Weitkamp2005}, the measured beam intensity can be expressed as follows
\begin{equation}
I^{(k)}=I_0+I_1\times\cos\left[2\pi\times\frac{x_{ps}}{\mathrm{p'}}+\phi_{obj}+\phi_{bkg}\right].
\label{eq:signal_mod}
\end{equation}
In it, $I_0$ is the mean intensity of the fringe, $I_1$ is the fringe amplitude, $\phi_{obj}$ corresponds to the object induced DPC signal, $\phi_{bkg}$ corresponds to the reference DPC signal without object, $x_{ps}$ is the PS length of the phase grating along $x$ axis, and $p'=p/n$ ($n$ is equal to 2 for $\pi$ phase grating, and is equal to 1 for $0.5\pi$ phase grating). According to Fig.~\ref{fig:fig2}, we have
\begin{align}
\label{eq:sens_1}
\phi_{obj} = 2\pi\times\frac{x_g}{p'},
\end{align}
where $x_g$ denotes the equivalent PS length of the phase grating corresponding to the $\Delta$ fringe shift, and is equal to
\begin{align}
\label{eq:sens_2}
x_g = \frac{d_1 d_3}{d_1+d_2+d_3}\varphi.
\end{align}
Substituting Eq.~(\ref{eq:sens_2}) into Eq.~(\ref{eq:sens_1}), and compare with Eq.~(\ref{eq:sens}), it is easy to demonstrate that the sensitivity $S$ is equal to
\begin{align}
\label{eq:sens_3}
S = 2\pi\times\frac{d_3}{p''}.
\end{align}
Herein, the $p''$ represents the effective self-imaging period modified by the fan-beam effect,
\begin{align}
\label{eq:sens_4}
p'' = p'\times\frac{d_1+d_2+d_3}{d_1},
\end{align}
The $d_1$ corresponds to the source-to-grating distance, $d_2$ corresponds to the grating-to-object distance, and $d_3$ corresponds to the object-to-image distance. When the object is positioned between the source and the grating, the sensitivity can be determined according to the principle of reversibility. It is
\begin{align}
\label{eq:sens_5}
S = 2\pi\times\frac{d_3}{p''}\times\frac{d_1}{d_1+d_2},
\end{align}
and
\begin{align}
\label{eq:sens_6}
p'' = p'\times\frac{d_1+d_2+d_3}{d_1+d_2}.
\end{align}
Herein, $d_1$ is the source-to-object distance, $d_2$ is the object-to-grating distance, and $d_3$ is the grating-to-fringe-image distance. When performing the above calculations, the corresponding distances should be altered accordingly, see Fig.~\ref{fig:fig2}.

So far, the sensitivity $S$ at any arbitrary position within an individual XOM can be predicted in theory immediately, as long as the imaging geometry and grating specifications are provided. Notice that this new theory does not need to know the period of the formed diffraction fringe image. Because every XOM works independently, as a result, this developed theory is suitable to perform sensitivity predictions for interferometers which having any number of XOMs (i.e., any number of phase gratings), as shown in Fig.~\ref{fig:fig1}(b)-(c). Because the sensitivity for single phase grating based Talbot-Lau interferometer system has been well studied\cite{Weitkamp2005, donath2009inverse}, this work mainly focuses on investigating the sensitivity of the dual-phase grating and the triple-phase grating interferometer systems. 

Experiments were performed on our benchtop to validate the developed sensitivity estimation theory, particularly for the dual-phase grating and the triple-phase grating interferometer systems. The X-ray imaging systems included a micro-focus X-ray tube (L9421-02, Hamamatsu Photonics, Japan) with $7.00~\mu \textrm{m}$ focal spot size. The micro-focus tube was operated at a tube voltage of 40.00 kV and a tube current of 190~$\mu$A. The X-ray CCD detector (OnSemi KAI-16000, XIMEA GmbH, Germany) has a native element dimension of 7.40~$\mu \textrm{m}$ and an effective imaging area of 36.00~mm$\times$24.00~mm. All the phase gratings had a period of 2.86~$\mu$m, with a duty cycle of 0.50. These gratings generated $\pi$ phase shifts for 17.00~keV X-ray photons. Their specifications were provided by the manufacturer (Microworks GmbH, Germany). The G1 grating was moved laterally by seven times with a step length of 0.40~$\mu$m. For each phase stepping position, the X-ray exposure period was 300 seconds. A homogeneous PMMA rod a diameter of 2.46~mm was imaged at a couple of different positions during the experiments.

With the acquired PS data, the standard signal retrieval method\cite{Momose2003, Weitkamp2005} was first implemented to obtain the DPC images. Then, the extracted DCP images were 10$\times$10 rebinned with the purpose to reduce signal noise. To determine the corresponding sensitivity at a certain position, the DPC signal was numerically simulated by inserting the PMMA rod phantom on the light path with the assumption of a point X-ray source (see the supplementary material for numerical simulation details). The goodness-of-fit between the experimental data and the numerical results was analyzed using the $R^2$ method.

\begin{figure}[t]
\centering
\includegraphics[width=0.8\textwidth]{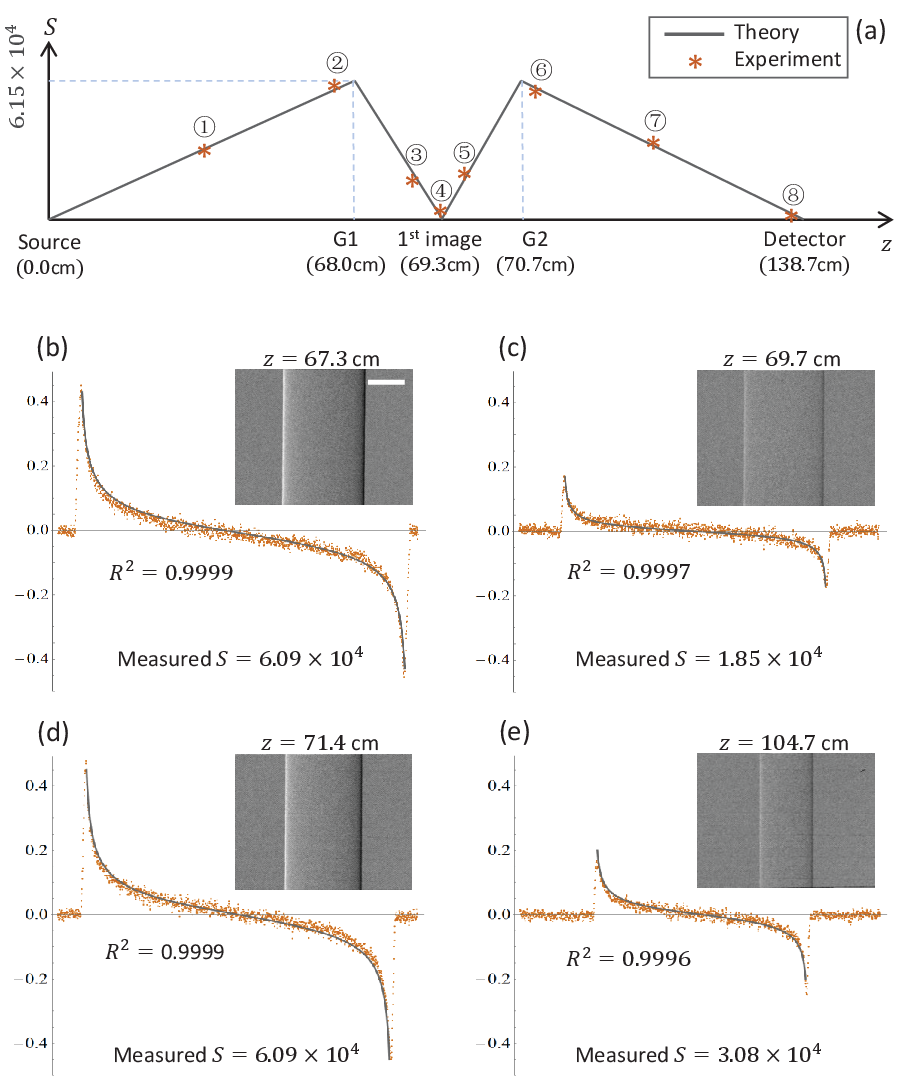}
\vspace{-0.1in}
\caption{The sensitivity response of a dual phase grating interferometer system. (a) Comparison results of theoretical predictions and the experimental measurements at 8 different positions. The DPC images, vertically averaged DPC signal profile (dotted line), and the theoretical DPC profile (solid line) for positions \textcircled{2}, \textcircled{5}-\textcircled{7}, are shown in (b)-(e), correspondingly. The scale bar denotes 5.0 mm.}
\label{fig:fig3}
\end{figure}

\begin{figure}[t]
\centering
\includegraphics[width=0.8\textwidth]{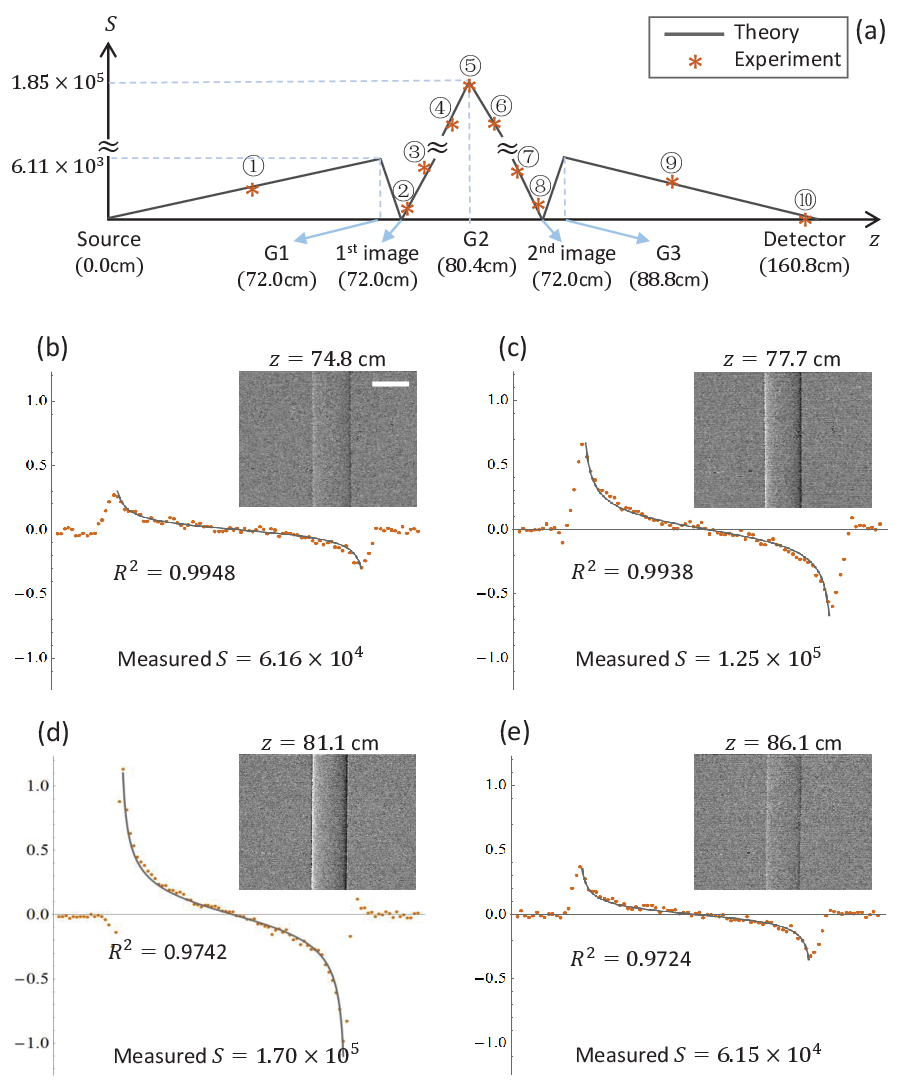}
\vspace{-0.1in}
\caption{The sensitivity response of a triple phase grating interferometer system. (a) Comparison results of theoretical predictions and the experimental measurements at 10 different positions. The DPC images, vertically averaged DPC signal profile (dotted line), and the theoretical DPC profile (solid line) for positions \textcircled{3}-\textcircled{5}, \textcircled{7}, are shown in (b)-(e), correspondingly. The scale bar denotes 5.0 mm.}
\label{fig:fig4}
\end{figure}

The sensitivity plot in Fig.~\ref{fig:fig3}(a) shows the strong consistency between the theory and the experimental measurements for the dual phase grating interferometer. As expected, the entire sensitivity curve between the X-ray source and the detector has the ``M'' shape. In particular, there are two sensitivity peaks with a valley ($S=0$) between them. Interestingly, the peaks appear exactly at the the phase grating planes, and the sensitivity valley appears right at the image plane of the X-ray source after the first phase grating. In addition, the sensitivity plot in Fig.~\ref{fig:fig4}(a) also shows the good consistency between the predictions and the experimental measurements for the triple phase grating interferometer. In this case, the entire sensitivity curve has three peaks and two valleys. Same as for the dual phase grating system, the sensitivity reaches to the peak at the phase grating planes, and decreases to the valley at the corresponding image planes. For our triple phase grating interferometer system, unfortunately, we did not observe the first and third sensitivity peaks because the sensitivities at these two locations are two orders less than the middle peak.

The PMMA rod DPC imaging results and the quantitative DPC signal profiles (both the measurements and numerical fittings) at different object positions are shown in Fig.~\ref{fig:fig3}(b)-(e) and Fig.~\ref{fig:fig4}(b)-(e). Overall, the measured DPC signals match well with the theoretical predictions, including both the shape and the value. Several factors may contribute errors to the numerical calculations, and thus leading to errors of the measured sensitivity $S$. For instance, the simulated X-ray beam spectrum, the estimated energy response of the interferometer, and the finite focal spot size. In this study, we put all these potential uncertainties together as measurement errors. Nevertheless, the measured sensitivity at different positions show strong consistency with the theoretical predictions, demonstrating the robustness of the developed theory for calculating interferometer's sensitivity, especially for the systems containing more than one phase gratings.

When using the dual $\pi$ phase grating system, it is recommended to put the object in front of the first grating or put it behind the second grating to avoid the sensitivity valley located between the two gratings. This is especially important if the object has a relatively large dimension. However, when using the triple $\pi$ phase grating system, it would be beneficial to put the object close to the second phase grating (either before it or behind it). For other triple phase grating setups, it might be possible to put the object before the first grating or behind the third grating (depending on the system design).

Both the theory and experiments show that the triple phase grating interferometer has better angular sensitivity performance, i.e., higher detected DPC signal values, than the dual phase grating interferometer, assuming the total system length and the phase grating pitches are similar. In addition, reducing the phase grating periods can help to boost the system sensitivity. Compared with the conventional Talbot-Lau interferometer, however, the less flexible system configurations may still impede the wide applications of the dual phase grating and triple phase grating interferometers. Therefore, careful system design and optimization are required.

For an individual XOM, we compared the theoretical sensitivity expressions with the previously published results\cite{Weitkamp2005, donath2009inverse}, and confirmed the validity of the derived sensitivity results for the assumed XOM of this work in a more general sense: First, our analyses are performed for the defined XOM in this work, instead of the Talbot-Lau interferometer; Second, the X-ray source in an XOM could be the diffraction fringe formed by the phase grating, rather than a real X-ray tube source; Third, the last component in XOM could be the diffraction fringe image, instead of the analyzer grating.

We have measured the sensitivity performance of the triple phase grating interferometer from experiments for the first time. Even though the imaging theory developed by Miao \emph{et. al.}\cite{Miao2016a} was employed to determine the system configurations, we realized that it might be possible to interpret such a special interferometer system by generalizing our previous theory\cite{ge2020dual} for the dual phase grating system. The main idea is illustrated in Fig.~\ref{fig:fig1}(c). Due to the space limit, rigorous theoretical analyses will be presented in another work in future.

In summary, this paper develops a novel framework to estimate the sensitivity of X-ray interferometers with multiple phase gratings. Validation experiments that are performed on interferometers with both two phase gratings and three phase gratings show good consistency between the predictions and the measurements. In future, this proposed approach can greatly facilitate the optimization and design of the multiple phase grating interferometer systems to achieve the best biomedical X-ray DPC imaging performance.

The authors would like to thank Dr. Zhicheng Li for providing the phase gratings. This project is supported by the National Natural Science Foundation of China (Grant No.~11804356, 11535015, 61671311).

The data that supports the findings of this study are available within the article [and its supplementary material].

\bibliography{Bibliography_Paper}% Produces the bibliography via BibTeX.

\end{document}